\documentstyle[12pt,psfig]{article}
\newcommand{\s}{\mathrm}
\newcommand{\bd}{\mathbf}
\newcommand{\ra}{\rightarrow}
\newcommand{\mn}{\mu \nu}
\newcommand{\be}{\begin{equation}}
\newcommand{\ee}{\end{equation}}
\newcommand{\ov}{\overline}
\newcommand{\ba}{\begin{eqnarray}}
\newcommand{\ea}{\end{eqnarray}}
\newcommand{\cm}{\cal{M}}
\newcommand{\ep}{\epsilon}
\newcommand{\lda}{\lambda}
\newcommand{\rpp}{\rho \pi \pi}
\newcommand{\opr}{\omega \pi \rho}

\begin{document}

\begin{flushright}
VECC/NTH 97013
\end{flushright}

\begin{center}
{\Large{Photons from Hadronic Matter at Finite Temperature }\\
}
\vskip .2in
Sourav Sarkar, Jan-e Alam, Pradip Roy\\

{\it Variable Energy Cyclotron Centre,
     1/AF Bidhan Nagar, Calcutta 700 064
     India}\\

Abhee K. Dutt-Mazumder, Binayak Dutta-Roy\\ 
{\it Saha Institute of Nuclear Physics,
           1/AF Bidhan Nagar, Calcutta 700 064
           India}\\

Bikash Sinha\\

{\it Variable Energy Cyclotron Centre,
     1/AF Bidhan Nagar, Calcutta 700 064
     India}\\
{\it Saha Institute of Nuclear Physics,
           1/AF Bidhan Nagar, Calcutta 700 064
           India}\\
\end{center}

\addtolength{\baselineskip}{0.5\baselineskip}
\parindent=20pt
\vskip 0.1in
\begin{abstract}
Temperature dependence of hadronic decay widths and masses
are studied within the framework of an effective Lagrangian approach.
At finite temperature the hadronic masses do not seem to follow 
a universal scaling law. Considering an exhaustive set of 
hadronic reactions and vector meson decays we have estimated the 
photon spectrum emitted from hot hadronic matter taking into account 
medium effects through thermal loop corrections on the hadronic decay widths
and masses. An enhancement in photon emission rate is obtained when we 
use the in-medium masses of vector mesons in our calculations.
It is observed that the effect of $\rho$ decay width on 
the photon spectra is negligible. 
\end{abstract}



\section*{I. Introduction}

   Investigation of photons emanating from hot and  dense hadronic matter 
formed in 
ultra-relativistic heavy-ion collisions is a field of considerable current
interest. As the temperature and/or density of hadronic matter
increases, it is expected that the system undergoes a phase transition
to a new state - quark-gluon plasma (QGP)~\cite{BM,Hwa,Wong}, in which
the quarks and gluons are locally free. 
Electromagnetic probes (photons and dileptons) have been proposed~\cite{Jane}
as one of the most promising signals of QGP formation; because of the 
very nature of their fundamental interactions, they tend to leave the 
system without much change in their initial energy and momentum,
thus, carrying the information of the reaction zone rather more 
effectively without being masked by the details of the evolution 
process, unlike what occurs for hadrons which interact strongly 
with the rest of the system.

    According to lattice QCD calculations, the critical temperature for the
phase transition is expected to be around 150 - 200 MeV. 
In an idealised first order phase transition scenario both hadronic 
matter and QGP would co-exist at the critical temperature until the 
cooling due to expansion dominates the heating due to liberation of 
latent heat.  The study of QCD phase transition in the laboratory is 
also important for understanding the evolution of the early universe 
a micro-second after the Big Bang~\cite{Bono}.
It is therefore essential to understand the hadronic properties 
at temperatures $\sim$ 150 -- 200 MeV. To clinch evidence for the
formation of QGP in its proper perspective it is essential 
to evaluate the rate of emission of photons from hadronic matter 
at finite temperature also, as they contribute to the background in 
the spectrum of photons from the hot matter formed.
For the present purpose we confine our attention to high 
temperature and zero net baryon density (and, therefore,
vanishing chemical potential). Thus our discussions relate to
future RHIC and LHC experiments rather than the present SPS data.
Inclusion of effects emanating from finite chemical potential
is relegated to a future communication.

    Irrespective of whether QGP is formed or not, hadronic matter
formed in ultra-relativistic heavy-ion collisions is expected to be
in a highly excited state. Hadronic interactions in a medium
at high temperature are not yet understood fully. There are several 
aspects where finite temperature medium effects may play an important 
role. For example, spontaneously broken chiral symmetry (a typical
characteristic of hadrons in their ground state) is expected to be restored
at high temperature~\cite{Pisarski}, which should manifest itself in 
the thermal shift of the hadronic masses as well as the change in their 
decay widths. These modifications of hadron properties e.g. mass, decay 
widths etc, can be studied through hadronic spectra (threshold for particle 
production depends on the mass of the particle) as well as from photon and 
dilepton spectra.

The low energy hadronic states require for their description, 
non-perturbative methods in QCD which are not available to us. 
However, the effective Lagrangian approach
is quite successful in explaining many of the observed properties
of hadronic interactions at the low energy scale~\cite{vol16}.
In this paper, by and large we follow the Quantum Hadrodynamics 
(QHD) model as discussed in Ref.~\cite{vol16}.

The properties of a meson get modified due to its interaction
with real and virtual excitations in the medium. 
In QHD, the scalar field is responsible for reduction 
of the nucleon mass. The vector meson mass falls due to the decrease of the
nucleon anti-nucleon masses, which appear through thermal loops in the
vector meson self energy. In an effective Lagrangian approach the change
in the vector meson mass comes from two different effects. On the one hand,
the scattering (space-like process~\cite{Jean}) of the probe with the 
real (on shell) particles present in the medium via vector meson exchange 
changes the mass of the vector meson. We observe
that such an effect brings in a small change in vector meson masses.
On the other hand, the effect of the infinite Dirac sea, which can
be probed by time-like processes, reduces the vector meson masses
once we take into account the medium modifications of the vacuum. 
Such an effect is included in our calculation by using the reduced
nucleon mass in the vacuum polarization. We observe that the vacuum
polarization overwhelms the corresponding medium dependent contribution.
In Ref.~\cite{Lee} the temperature dependence of the vector meson 
masses has been studied by using QCD sum rules. According to these calculations 
the vector meson mass falls with temperature due to modification of 
the vacuum at finite temperature. In this approach the decrease 
in the vector meson masses results from the decrease of the quark condensate
with temperature. 

The change in the vector meson masses and decay widths modifies
the photon production cross section as well as the available
phase space in a thermalised system. We evaluate the photon
spectra from  hot hadronic matter with and without temperature
dependent masses and decay widths. It is observed that 
although the effect of temperature dependent decay
width is negligible, the effect of the temperature dependent masses 
is important in the photon emission rate.

   We organise the paper as follows. In section II we calculate temperature 
dependent properties of the rho and omega mesons
within the framework of an effective Lagrangian approach. 
In section III we present the calculation of photon rates 
from hot hadronic matter. Section IVa will be utilised to discuss 
finite temperature effects on hadronic masses and their decay widths. 
Section IVb will be devoted to the study of the effects of temperature 
dependent masses and decay widths on photon production rates and in 
section V, we present a summary and conclusions.

\section*{II. Finite Temperature Effects}

     To  study and understand hadronic matter produced in ultra-relativistic 
heavy-ion collisions at high temperature, finite temperature field theory 
is the most consistent approach. The propagation of a particle 
(say $\rho$ or $\omega$) in a heat bath is modified as its pole is shifted, 
thereby changing the mass of the particle. In the present calculation thermal 
effects enter through thermal nucleon and pion loops. 

In the following, we briefly discuss the effective (dressed) 
propagators taking into account the in-medium effects.
We begin with the Dyson equation:  

\begin{equation}
D_{\mu \nu}^{-1} = (D_{\mu \nu}^{0})^{-1} - \Pi_{\mu \nu},
\label{dyson}
\end{equation}
where 
\begin{equation}
D_{\mu \nu}^{0} = \frac{-g_{\mu \nu}+k_{\mu}k_{\nu}/m_V^2}
{k^2-m_V^2+i\epsilon}
\label{freeprop}
\end{equation}
is the free-space (bare) propagator; $k^{\mu} = (\omega, {\mathbf k})$
is the four-momentum of the propagating particle and $m_V$ is the mass of 
the vector meson. $\Pi_{\mu \nu}$ is the self energy of the vector meson 
given by,
\begin{equation}
\Pi^{\mu \nu}=\Pi_{\s {vac}}^{\mu \nu}+\Pi_{\s {med}}^{\mu \nu},
\label{pitot}
\end{equation}
where
\begin{equation}
\Pi_{\s {vac}}^{\mu \nu}=(g^{\mu \nu} - \frac{k^{\mu}k^{\nu}}{k^2})\,
\Pi_{\s {vac}}(k^2) 
\label{pivac}
\end{equation}
is the Dirac sea contribution to the self energy \cite{hatplb}. In a thermal 
bath moving with four-velocity $u^{\mu}$, $\Pi_{\s {med}}^{\mu \nu}$
has transverse and longitudinal components~\cite{Adas},

\begin{equation}
\Pi_{\s {med}}^{\mu \nu}(\omega,{\mathbf k}) = A^{\mu \nu}\Pi_{T,{\s {med}}} +
 B^{\mu \nu}\Pi_{L,{\s {med}}}.
\label{pimed}
\end{equation}
$A^{\mu \nu}$ and $ B^{\mu \nu}$ are the transverse and longitudinal
projection tensors given by

\be
A^{\mn}=\frac{1}{k^2-\omega^2}\left[(k^2-\omega^2)(g^{\mn}-u^{\mu}u^{\nu})
\frac{}{}-k^{\mu}k^{\nu}-\omega^2u^{\mu}u^{\nu}+\omega(u^{\mu}k^{\nu}+
k^{\mu}u^{\nu})\right]
\ee
and,
\be
B^{\mn}=\frac{1}{k^2(k^2-\omega^2)}\left[\frac{}{}\omega^2k^{\mu}k^{\nu}+
k^4u^{\mu}u^{\nu}-\omega k^2(u^{\mu}k^{\nu}+k^{\mu}u^{\nu})\right],
\ee
which obey the relation

\begin{equation}
A^{\mu \nu} + B^{\mu \nu} = g^{\mu \nu} - \frac{k^{\mu}k^{\nu}}{k^2}.
\label{ab}
\end{equation}
Using Eqs.~(\ref{dyson}-\ref{ab}) the effective propagator becomes

\begin{equation}
D_{\mu \nu} = -\,\frac{A_{\mu \nu}}{k^2-m_V^2+\Pi_{T}}
-\,\frac{B_{\mu \nu}}{k^2-m_V^2+\Pi_{L}} + \frac{k_{\mu}k_{\nu}}{k^2\,m_V^2},
\label{deff}
\end{equation}
Here, 
\begin{equation}
\Pi_{T(L)}=\Pi_{T(L),{\s {med}}}+\Pi_{\s {vac}}.
\end{equation}

Thus we see from Eq.~(\ref{deff}) that the diagram defining the self energy 
$\Pi^{\mu \nu}$ in a thermal bath shifts the pole of the propagator.
The real part of the self energy is responsible for mass shifting and
the imaginary part gives the decay width of the vector meson in the medium. 
The effective mass of the vector meson in the medium can be obtained from 
the pole position of the propagator in the limit ${\mathbf k}\,\rightarrow 0$
{\it i. e.} in the rest frame of the vector meson. In this limit 
$\Pi_{T,{\s {med}}} = \Pi_{L,{\s {med}}} = \Pi_{\s {med}}$, and we have,
\be
\frac{1}{3}\Pi_{\mu}^{\mu}=\Pi= \Pi_{\s {med}}+\Pi_{\s {vac}}
\ee   
The effective mass of the vector meson is then obtained by solving the 
equation:

\begin{equation}
\omega^2 - m_V^2 + {\mathrm {Re}}\Pi = 0
\label{mass}
\end{equation}
 
\subsection*{IIa. $VNN$ Interaction}

\begin{figure}
\centerline{\psfig{figure=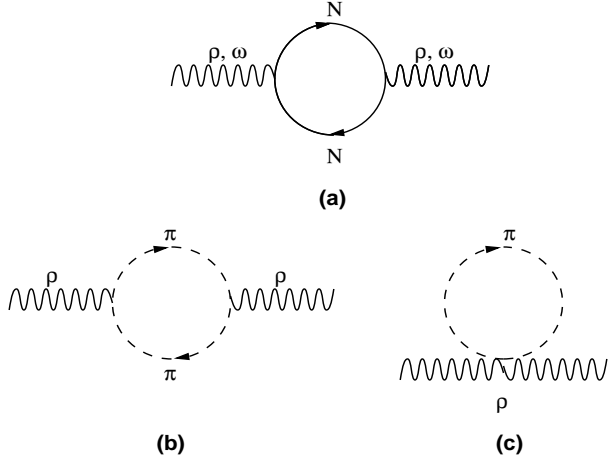,height=6cm,width=8cm}}
\caption{
Feynman Diagrams for self energy of vector mesons}
\label{fdself}
\end{figure}

To calculate the effective mass and decay widths we begin with the 
following $VNN$ (Vector - Nucleon - Nucleon) Lagrangian density:

\begin{equation}
{\cal L}_{VNN} = g_{VNN}\,\left({\bar N}\gamma_{\mu}
\tau^a N{V}_{a}^{\mu} - \frac{\kappa_V}{2M}{\bar N}
\sigma_{\mu \nu}\tau^a N\partial^{\nu}V_{a}^{\mu}\right),
\label{lag1}
\end{equation}
where $V_a^{\mu} = \{\omega^{\mu},{\vec {\rho}}^{\mu}\}$,
$M$ is the free nucleon mass, $N$ is the nucleon field
and $\tau_a=\{1,{\vec {\tau}}\}$.
The values of the coupling constants $g_{VNN}$ and $\kappa_V$ will 
be specified later.  With the above Lagrangian we proceed to calculate the 
$\rho$-self energy. The relevant Feynman diagram is shown in Fig.~(1a). 
The one-loop correlation function is given by,

\begin{equation}
\Pi_{\mu \nu} = -2ig_{VNN}^2\,\int\,
\frac{d^4{p}}{(2\pi)^4}\,S_{\mu \nu}(p,k),
\label{pimunu}
\end{equation}
with,
\begin{equation}
S_{\mn} = \frac{{\mathrm {Tr}}\left[\Gamma_{\mu}(k)\,(p\!\!\!/+M^{\ast})
\Gamma_{\nu}(-k)\,(p\!\!\!/-k\!\!\!/+M^{\ast})\right]}{(p^2-M^{\ast 2})
((p-k)^2-M^{\ast 2})},
\label{smn}
\end{equation}
The vertex $\Gamma_{\mu}(k)$ is calculated by using the
Lagrangian of Eq.~(\ref{lag1}) and is given by

\begin{equation}
\Gamma_{\mu}(k) = \gamma_{\mu} + \frac{i\kappa_V}{2M}\sigma_{\mu \alpha}
k^{\alpha}.
\label{vertex}
\end{equation}
Here $M^{\ast}$ is the in-medium (effective) mass of the nucleon at finite 
temperature which we calculate using the Mean-Field Theory (MFT)~\cite{vol16}.
The value of $M^{\ast}$ can be found by solving the following 
self consistent equation:

\begin{equation}
M^{\ast} = M -\frac{4g_s^2}{m_s^2}\,\int\,\frac{d^3{\mathbf p}}{(2\pi)^3}\,
\frac{M^{\ast}}{({\mathbf p}^2+M^{\ast 2})^{1/2}}\,\left[f_N(T)+f_{\bar N}(T)\,\right],
\label{MN}
\end{equation}
where $f_N(T)(f_{\bar N}(T))$ is the Fermi-Dirac distribution for 
the nucleon (antinucleon), $m_s$ is mass of the neutral scalar meson
($\sigma$) field, and, the nucleon interacts via the exchange of 
isoscalar meson with coupling constant $g_s$.  Since the exact 
solution of the field equations in QHD is untenable, these are solved in 
the mean field approximation. In a mean field approximation one
replaces the field operators by their ground state expectation
values which are classical quantities~\cite{vol16}; this 
renders the field equations exactly solvable.

The vacuum part of the $\rho$ self energy arises due to its interaction 
with the nucleons in the modified Dirac sea. This is calculated using 
dimensional regularization scheme to yield,
\begin{equation}
{\s {Re}}\Pi_{\s {vac}}(\omega, {\bd k} \ra \, 0)=-\frac{g_{VNN}^2}{\pi^2}
\omega^2\,\left[I_1+M^{\ast}\frac{\kappa_V}{2M}I_2+\frac{1}{2}\,(\frac{\kappa_V}
{2M})^2\,(\omega^2I_1+M^{\ast 2}I_2)\right]
\label{pik2}
\end{equation}
where,
\be
I_1=\int_{0}^{1}\,dz\,z(1-z)\,\ln\left[\frac{M^{\ast 2}-\omega^2\,z(1-z)}
{M^2-m_V^2\,z(1-z)}\right],
\ee
\be
I_2=\int_{0}^{1}\,dz\,\ln\left[\frac{M^{\ast 2}-\omega^2\,z(1-z)}
{M^2-m_V^2\,z(1-z)}\right].
\ee

Here, it is important to note that we have used the in-medium nucleon
mass in the vacuum part of the self energy of the vector meson.

In a hot system of particles, there is a thermal distribution of real particles
(on shell) which can participate in the absorption and emission process in
addition to the exchange of virtual particles. The interaction of the rho 
with the onshell nucleons, present in the thermal bath contributes to the 
medium dependent part of the $\rho$-self energy.  This is calculated from
Eq.(\ref{pimunu}) using imaginary time formalism~\cite{Morley,Kapusta} as,

\begin{eqnarray}
{\s {Re}}\Pi_{\s {med}}(\omega,{\bd k}\,\ra\,0)& = &
-\frac{8g_{VNN}^2}{\pi^2}\,\int\,\frac{p^2\,dp}{\omega_p\,
(e^{\beta\,\omega_p}+1)(4\omega_p^2-\omega^2)}\nonumber\\
&&\times\,\left[\frac{2}{3}(2p^2+3M^{\ast 2})-\omega^2\left\{2M^{\ast}
(\frac{\kappa_V}
{2M})\right.\right.\nonumber\\
&&-\left.\left.\,\frac{2}{3}(\frac{\kappa_V}{2M})^2(p^2+3M^{\ast 2})\right\}
\right]
\end{eqnarray}
where, $\omega_p^2=p^2+M^{\ast 2}$.
Note that in the above equation  the sign of the term proportional
to $\kappa_V$ is opposite to that obtained in Ref.~\cite{Song} 
(see Eq.~(14) of Ref.~\cite{Song}).

\subsection*{IIb. $\rho \pi \pi$ Interaction}

The change in the properties of $\rho$ meson due to rho pion coupling
has been studied in a similar manner. The effect in this case is 
small~\cite{Gale} compared to the nucleon loop contribution. This is 
attributed to the fact that the pion mass does not change in the medium 
but the nucleon mass drops substantially due to mean-field effects. 
To evaluate the rho self-energy due to $\pi\pi$ loop we use the following 
well known interaction Lagrangian: 
\begin{equation}
{\cal L}_{\rho \pi \pi} = -g_{\rho \pi \pi}\,{\vec {\rho}^{\mu}}\cdot
({\vec {\pi}}\, \times\,\partial_{\mu}{\vec {\pi}})+
\frac{1}{2}g_{\rho \pi \pi}^2(\vec {\rho}^{\mu}\times\vec {\pi})\cdot
(\vec {\rho}_{\mu}\times\vec {\pi})
\label{lag2}
\end{equation}
where the vector signs on the $\rho$ and $\pi$ indicate that they are
isovectors.

In this case the diagrams 1(b)and 1(c) contribute to the rho self energy in the
hot medium and is given by,
\be
\Pi^{\mn}=ig_{\rho \pi \pi}^2\,\int\frac{d^4p}{(2\pi)^4}\,
\frac{(2p+k)^{\mu}(2p+k)^{\nu}}{[(p+k)^2-m_{\pi}^2][p^2-m_{\pi}^2]}
-2ig_{\rho \pi \pi}^2\,\int\,\frac{d^4p}{(2\pi)^4}\,\frac{1}{[p^2-m_{\pi}^2]}
\label{pipimn}
\ee
This is again evaluated using the imaginary time formalism to give,
\begin{equation}
{\s {Re}}\Pi(\omega,{\mathbf k}\,\rightarrow\,0) = -\frac{g_{\rho \pi \pi}^2}
{3\pi^2}\,\int\,\frac{p^2\,dp}{\omega_p}\,\left[3 - \frac{4p^2}{4\omega_p^2-
\omega^2}\,\right]\frac{1}{e^{\beta \omega_p}-1}
\label{piloop}
\end{equation}
Here, $\omega_p^2=p^2+m_{\pi}^2$.
We then calculate the change in
$\rho$-mass due to pion loop using Eqs.~(\ref{mass}) and (\ref{piloop}).

  The decay width for the process $\rho\,\rightarrow\,\pi\,\pi$ can be
calculated from the imaginary part of the self energy 
using cutting rules \cite{Weldon} which, in the rest
frame of $\rho$-meson is given by (see Appendix - I)
\begin{equation}
\Gamma_{\rho\,\rightarrow\,\pi\,\pi} = \frac{g_{\rho}^2}{48\pi\,\omega^2}\,
(\omega^2-4m_{\pi}^2)^{3/2}
\left[\left(1+n(\frac{\omega}{2})\right)\,\left(1+n(\frac{
\omega}{2})\right)-n(\frac{\omega}{2})n(\frac{\omega}{2})
\right]
\label{width}
\end{equation}
where $\omega = m_{\rho}^{\ast}$ is the in-medium mass of $\rho$-meson. 
$n(\omega/2)$ is Bose-Einstien distribution for the $\pi$-meson. The 
first term in Eq.(\ref{width}) represents the stimulated emission of 
pions with statistical weight $(1+n(\omega /2))(1+n(\omega /2))$ and 
the second term stands for the absorption of pions with weight factor 
$n(\omega /2)n(\omega /2)$.

It has been emphasized recently~\cite{Abhee} that the medium effects
on the decay width of the $\rho$-meson play a very important role.
The thermal nucleon loop brings about a modification of the $\rho$-dominated
pion form factor which is further influenced by the thermal distributions
of the pions at a  finite temperature. The medium effects on dilepton spectra 
through the decay width of $\rho$-meson has been shown to be very important in
Ref.~\cite{Abhee}. In our calculation of photon spectra we include such 
effects through the rho-propagator.

\section*{III. Photons from Hot Hadronic Matter}

Whether a QCD phase transition takes place or not, photons can be used 
as a probe to study the properties of hadrons in hot/dense medium. 
In a phase transition scenario, apart from the QGP, photons are also 
produced from the thermalised hadronic gas, formed after the phase transition. 
Substantial contributions to the total photon yield also come from
the initial hard collision of partons in the high momentum
regime, and from hadronic decays ($\pi^{0}\,\rightarrow\,\gamma\,\gamma, 
\eta\,\rightarrow\,\gamma\,\gamma$ etc.) in the low momentum zone.
The hard QCD photons can be well understood through perturbative QCD 
and the decay photons can be reconstructed by invariant mass analysis. 
Thus, to extract photon signals from QGP it is essential to estimate 
the photon rates from various hadronic reactions and vector meson decays 
which is a challenging task, indeed. The temperature of the hadronic phase 
lies between 150 - 200 MeV. Therefore the finite temperature corrections 
to the hadronic properties and their consequences on photon spectra  
are very important.

Many of the aspects of the interacting pion gas have been reported
in the literature~\cite{Rapp,Chanfray,Bertsch}. For our purpose
we model the hadronic gas as consisting of $\pi, \rho, \omega$ and
$\eta$. First we consider the reactions $\pi\,\pi\,\rightarrow\,\rho\,\gamma$, 
$\pi\,\rho\,\rightarrow\,\pi\,\gamma$ and the decay $\rho\ra \pi \pi\gamma$. 
We estimate the differential cross-section for photon production from 
the above processes taking into account the finite width of the rho 
meson (see Appendix - II). The relevant vertices are obtained from 
the following Lagrangian:

\begin{equation}
{\cal L} = -g_{\rho \pi \pi}{\vec {\rho}}^{\mu}\cdot
({\vec \pi}\times\partial_{\mu}{\vec \pi}) - eJ^{\mu}A_{\mu} + \frac{e}{2}
F^{\mu \nu}\,({\vec \rho}_{\mu}\,\times\,{\vec \rho}_{\nu})_3,
\label{photlag}
\end{equation}
where $F_{\mu \nu} = \partial_{\mu}A_{\nu}-\partial_{\nu}A_{\mu}$, is the
Maxwell field tensor and $J^{\mu}$ is the hadronic part of the electromagnetic
current given by
\begin{equation}
J^{\mu} = ({\vec \rho}_{\nu}\times{\vec B^{\nu \mu}})_3 + (
{\vec \pi}\times(\partial^{\mu}\pi+g_{\rho \pi \pi}{\vec \pi}\times{\vec 
\rho}^{\mu}))_3
\label{jmu}
\end{equation}
with ${\vec B_{\mu \nu}} = \partial_{\mu}{\vec \rho}_{\nu}-\partial_{\nu}
{\vec \rho}_{\mu}-g_{\rho \pi \pi}(\vec \rho_{\mu}\times\vec \rho_{\nu})$,
and the subscript 3 after the cross product indicates the relevant component
in isospin space. The last term in Eq.~(\ref{photlag}) as well as
the first term in Eq.~(\ref{jmu}) arise due to the
non-abelian structure of the $SU(2)\,\times\,U(1)$ gauge group~\cite{vol16}. 
The 2$\rho$-1$\gamma$ vertex has been calculated by using the last
two terms of the lagrangian (\ref{photlag}) to obtain  gauge invariant
amplitudes.

   We have also considered the photon production due to the reactions 
$\pi\,\eta\,\rightarrow\,\pi\,\gamma$, $\pi\,\pi\,\rightarrow\,\eta\,
\gamma$ and the decay $\omega\,\ra\,\pi\,\gamma$ using the following 
Lagrangian~\cite{Lichard,JK}:
\be
{\cal L} = 
\frac{g_{\rho \rho \eta}}{m_{\eta}}\,
\epsilon_{\mu \nu \alpha \beta}\partial^{\mu}{\rho}^{\nu}\partial^{\alpha}
\rho^{\beta}\eta
+\frac{g_{\omega \rho \pi}}{m_{\pi}}\,
\epsilon_{\mu \nu \alpha \beta}\partial^{\mu}{\omega}^{\nu}\partial^{\alpha}
\rho^{\beta}\pi^0
+\frac{em_{\rho}^2}{g_{\rho \pi \pi}}A_{\mu}\rho^{\mu}
\label{etaro}
\ee 
The last term in the above Lagrangian is written down on the basis
of Vector Meson Dominance (VMD) ~\cite{Sakurai}.

In the case of nuclear collisions, one is more interested in overall
photon rates rather than the cross-sections. 
The rate of emission of thermal photons with energy $E$ and
momentum $p$ is given by
\be
E\frac{dR}{d^3p}=-\frac{2g^{\mn}}{(2\pi)^3}{\s {Im}}\Pi_{\mn}^R\,(p)\frac{1}
{e^{\beta E}-1}
\label{photrate1}
\ee
where ${\s {Im}}\Pi_{\mn}^R$ is the imaginary part of the
retarded photon self energy 
at finite temperature. The emission rate given by Eq.~(\ref{photrate1}) 
is correct upto order $e^2$ in electromagnetic interaction but exact
(albeit in principle) to all orders in strong interaction. 
If ${\s {Im}}\Pi_{\mn}^R$ is evaluated
by carrying out loop expansion up to a finite order and applying
the cutting rules ({\it \'{a} l\'{a}} finite temperature modification of the
Cutkosky-Landau prescriptions)~\cite{Weldon} then the photon
emission rate becomes equivalent to the 
one obtained in relativistic kinetic theory formalism~\cite{Gale}. 
It may be worthwhile to mention that for a typical photon producing reaction 
($x$ particles $\ra$ $y$ particles $+ \gamma$), the above two formalisms
will give equivalent results if the photon self-energy is approximated
by carrying out a loop expansion upto and including $L$ loops where
$L$ satifies $x+y < L+1$~\cite{Gale}.

In this article we compute the rate  
of photon production per unit volume at a temperature $T$ using the
relativistic kinetic theory approach~\cite{Gal},
\begin{eqnarray}
E\frac{dR}{d^3p}&=&\frac{\cal N}{16(2\pi)^7E}\,\int_{(m_1+m_2)^2}^{\infty}
\,ds\,\int_{t_{\s {min}}}^{t_{\s {max}}}\,dt\,|{\cal M}|^2\,
\int\,dE_1\nonumber\\
&&\times\int\,dE_2\frac{f(E_2)\,f(E_2)\left[1+f(E_3)\right]}{\sqrt{aE_2^2+
2bE_2+c}}
\label{photrate}
\end{eqnarray}
where ${\cal M}$ is the invariant amplitude (see Appendix II)
of photon production for the appropriate reaction channel, evaluated 
from the Lagrangians given by Eqs. (\ref{photlag}) and (\ref{etaro}).
It may be emphasized that here, as elsewhere in the literature,
unstable external particles are taken to be on-shell with medium
modified masses. For convenience we show the parameters $a, b, c$ 
and also the limits of the integrals in Appendix III.

\section*{IV. Results}

\subsection*{IVa. Finite Temperature Properties}

\begin{figure}
\centerline{\psfig{figure=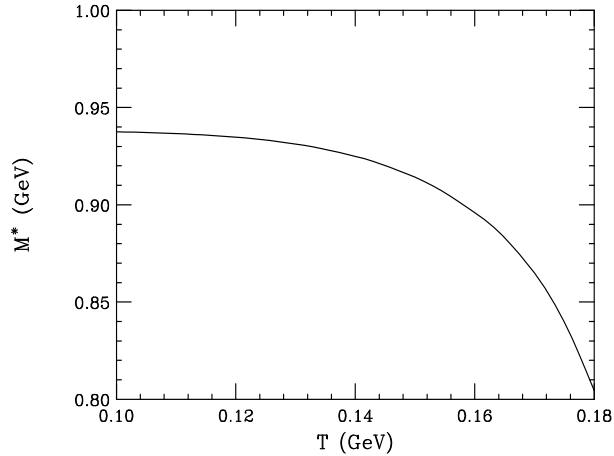,height=6cm,width=8cm}}
\caption{
Effective nucleon mass ($M^{\ast}$) as a function of temperature ($T$).
}
\label{nmass}
\end{figure}

   In this section we present the results of our calculation of the effective
masses and decay widths of vector mesons. For our calculations of $\rho$-meson 
effective mass we have used the following values of the coupling constants 
and masses~\cite{Bonn}: $\kappa_{\rho} = 6.1,~g_{\rho NN}^2/4\pi = 0.55, 
m_s$= 550 MeV, $m_{\rho} = 770$ MeV, $M = 938$ MeV, and $g_s^2/4\pi = 9.3$. 
The variation of nucleon mass due to the mean field plays a vital
role in determining the effective mass of $\rho$ and $\omega$ mesons.
The effective mass of the nucleon is calculated from Eq.~(\ref{MN}) and
its variation with temperature is plotted in Fig.~(\ref{nmass}). 
It is seen that the nucleon mass remains almost constant upto
a temperature of 140 MeV beyond which it starts falling.
We have plotted our results upto a temperature of 180 MeV above which
the effective nucleon mass decreases very sharply.

\begin{figure}
\centerline{\psfig{figure=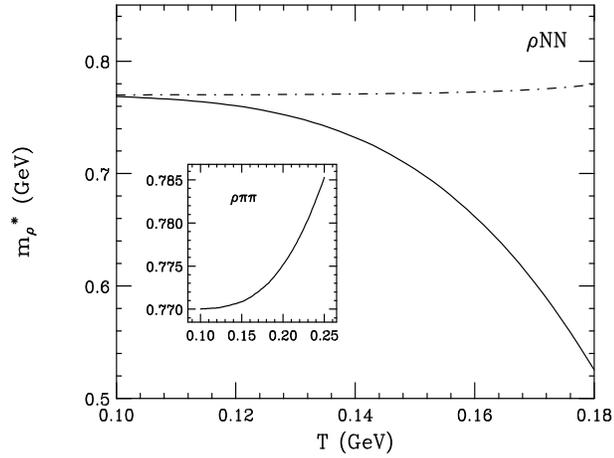,height=6cm,width=8cm}}
\caption{
Effective rho mass ($m_{\rho}^{\ast}$) as a function of temperature ($T$).
The dot-dashed line shows the effective mass due to $N{\bar N}$ polarisation
at finite temperature whereas the solid line includes the effect of fluctuations
of the Dirac sea of nucleons with mass $M^{\ast}$. The inset shows the
effective rho mass due to $\pi \pi$ polarisation at finite temperature.
}
\label{rhomass}
\end{figure}

We now consider the effect of $N{\bar N}$ polarisation on the 
$\rho$-meson mass. The vacuum and in-medium contributions are
shown separately in Fig.~(\ref{rhomass}). We observe that
the in-medium contribution increases very little ($\sim$ 5 -- 10 MeV) 
with temperature from its free space value. This is similar to the result
reported by Song {\it et. al}~\cite{Song}.
The overall decrease ($\approx 32\%$ at $T$ = 180 MeV) in the effective 
mass can be attributed almost entirely to fluctuations in the Dirac sea of
nucleons with mass $M^{\ast}$. The effective mass of the $\rho$ due to 
$\pi\pi$ polarisation is also shown in the inset of Fig.~(\ref{rhomass}).
In this case we observe a very small ($2\%$) increase in the effective 
mass at $T=$180 MeV.

\begin{figure}
\centerline{\psfig{figure=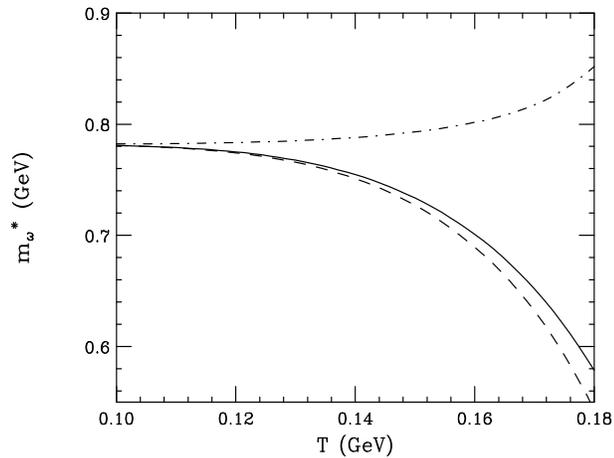,height=6cm,width=8cm}}
\caption{
Effective omega mass ($m_{\omega}^{\ast}$) as a function of temperature ($T$).
The dot-dashed line shows the effective mass due to $N{\bar N}$ polarisation
at finite temperature and the dashed line shows the effect 
of fluctuations
of the Dirac sea of nucleons with mass $M^{\ast}$. Solid line represents
the net effect.
}
\label{ommass}
\end{figure}

To calculate the effective mass of $\omega$-meson, we take 
$g_{\omega NN}^2/4\pi = 20$, $\kappa_{\omega}$ = 0~\cite{Bonn}. The result is
shown in Fig.~(\ref{ommass}). We observe the same feature as 
Fig.~(\ref{rhomass}) with an increase of about 70 MeV at $T=$ 180 MeV
in the in-medium contribution. The overall decrease in omega mass 
at $T=$180 MeV is about 200 MeV from its free space value.

\begin{figure}
\centerline{\psfig{figure=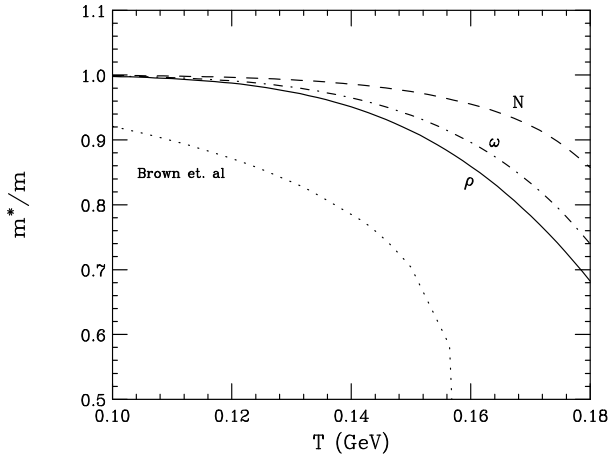,height=6cm,width=8cm}}
\caption{
Ratio of effective mass to free space mass of hadrons as a 
function of temperature
$T$.}
\label{scaling}
\end{figure}

Many authors have investigated the issue of temperature dependence of
hadronic masses within different models over the past several years.
Hatsuda and collaborators~\cite{Furn,Adami,Hatsuda} and Brown~\cite{Brown}
showed that the use of QCD sum rules at finite temperature results in
a temperature dependence of the $q \bar q$ condensate culminating
in the following behaviour of the $\rho$-mass:
\be
\frac{m_{\rho}^{\ast}}{m_{\rho}} = \left(1- \frac{T^2}{T_{\chi}^ 2}\right)^
{1/6}
\ee
where $T_{\chi}$ is the critical temperature for chiral phase
transition. Brown and Rho~\cite{Rho} also showed that the requirement 
of chiral symmetry yields an approximate scaling relation between 
various effective hadron masses,
\be
\frac{M^{\ast}}{M}\approx\frac{m_{\rho}^{\ast}}{m_{\rho}}\approx
\frac{m_{\omega}^{\ast}}{m_{\omega}}\approx\frac{f_{\pi}^{\ast}}{f_{\pi}}
\ee
These calculations show a dropping of hadronic mass with temperature. 
Calculations with non-linear $\sigma$-model, however,  
predict the opposite trend \cite{abijit}. In order to make a 
comparative study we plot the ratios of the effective masses to the free masses
of $\rho, \omega$ and $N$ versus temperature in Fig.~(\ref{scaling}). The
expected trend based on Brown - Rho scaling is also shown. 
Firstly, we do not observe any global scaling behaviour. Secondly, 
in our case the effective mass as a function of temperature falls
at a slower rate.  

\begin{figure}
\centerline{\psfig{figure=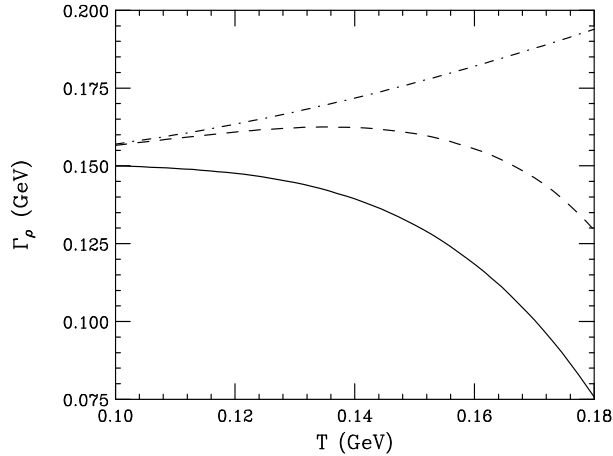,height=6cm,width=8cm}}
\caption{
Rho decay width ($\Gamma_{\rho}$) as a function of temperature ($T$).
Dashed and solid lines show calculations of rho width with effective mass
due to nucleon loop, with and without BE. Dot dashed line represents
the same but with effective mass due to pion loop. 
}
\label{rowidth}
\end{figure}

The variation of the in-medium decay width ($\Gamma_{\rho}$) of the rho meson
with temperature is shown in Fig.~(\ref{rowidth}). As discussed earlier,
the effective mass of the rho decreases as a result of $N\bar N$ 
polarisation. This reduces the phase space available for the rho. Hence, 
we observe a rapid decrease in the rho meson width with temperature 
(solid line). However, the presence of pions in the medium would
cause an enhancement of the decay width through induced emission. Thus
when Bose Enhancement (BE) of the pions is taken into account 
the rho decay width is seen to fall less rapidly (dashed line);
such a behaviour is observed quite clearly in ~\cite{Abhee}.
For the sake of completeness, we also show the variation of rho width in
the case where the rho mass changes due to $\pi \pi$ loop.
In this case, since the rho mass increases (though only marginally), the width 
increases (dot-dashed line).

\section*{IVb. Photon Spectra}

\begin{figure}
\centerline{\psfig{figure=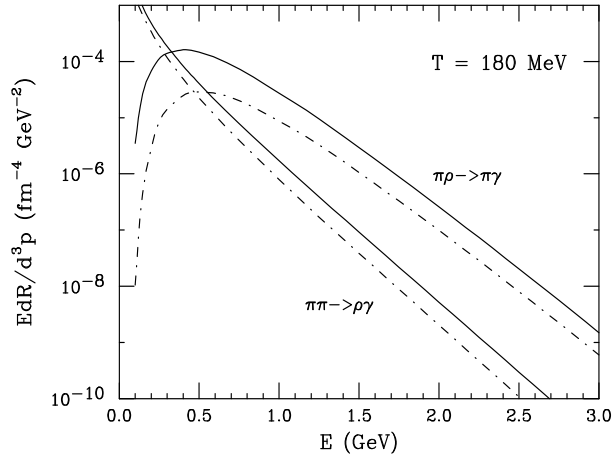,height=6cm,width=8cm}}
\caption{
Photon emission rates from $\pi \pi \ra \rho \gamma$ and $\pi \rho \ra
\pi \gamma$ as a function of photon energy at $T$=180 MeV. 
The solid and dot-dashed lines show results
with and without in-medium effects respectively.
}
\label{phot18}
\end{figure}

In this section we present our results on photon emission rates from a
hot hadronic gas. As discussed earlier, the variation of hadronic decay
widths and masses will affect the photon spectra. The relevant reactions
of photon production are $\pi \pi\,\ra\,\rho\,\gamma, \pi\,\rho\,\ra
\pi\,\gamma, \pi\,\pi\,\ra\,\eta\,\gamma$, and $\pi\,\eta\,\ra\,\eta\,\gamma$
with all possible isospin combinations. Since the lifetimes of the rho and
omega mesons are comparable to the strong interaction time scales,
the decays $\rho\,\ra\,\pi\,\pi\,\gamma$ and $\omega\,\ra\,\pi^0\,\gamma$
are also included. Contributions from $\rho\,\ra\,\pi\,\gamma$ and 
$\omega\,\ra\,\pi\,\pi\,\gamma$  are ignored because of very small
decay widths of these processes. We will consider the first two reactions 
in detail since they are seen to dominate the total photon yield. 
The effect of finite resonance width of the rho meson in the photon 
production cross-sections has been taken into account through the
propagator, which was neglected in Ref.~\cite{JK}. 

\begin{figure}
\centerline{\psfig{figure=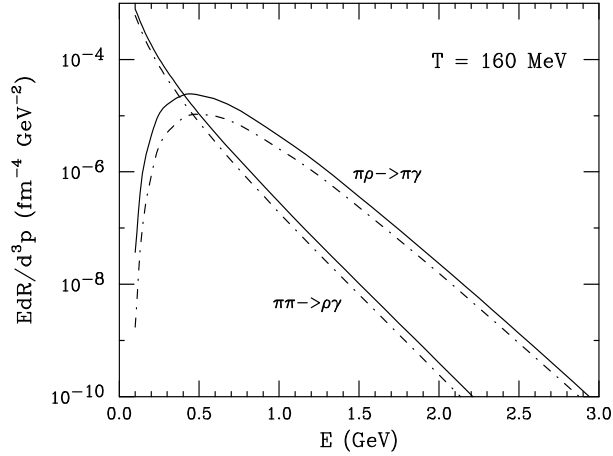,height=6cm,width=8cm}}
\caption{
Photon emission rates from $\pi \pi \ra \rho \gamma$ and $\pi \rho \ra
\pi \gamma$ as a function of photon energy at $T$=160 MeV. 
The solid and dot-dashed lines show results
with and without in-medium effects respectively.
}
\label{phot16}
\end{figure}

  In Fig.~(\ref{phot18}) the photon emission rates from the first two 
reactions ($\pi\,\pi\,\ra\,\rho\,\gamma$ and $\pi\,\rho\,\ra\,\pi\,\gamma$)
with and without finite temperature effects are presented at $T$ = 180 MeV. 
When in-medium masses and decay widths are taken into account, 
the photon yields from both the reactions are found to increase compared 
to the case when free masses and widths are considered. 
To understand this, let us recall that the emission rate of photons 
from a reaction $h_1\,h_2\,\ra\,h_3\,\gamma$ can  be approximated 
as~\cite{Wong} 
\begin{eqnarray}
E\frac{dR}{d^3p} &\propto&f_{h_1}(p_{\gamma})\,\int\,ds\,\frac{dE_{h_2}}
{E_{\gamma}}\,f_{h_2}(1+f_{h_3})\nonumber\\
&&\times\,\sqrt{s(s-4m^2)}\,\sigma_{h_1h_2\,\ra\,h_3\,\gamma}
\end{eqnarray}
where $m$ is the mass of the particles in the incident channel 
$(=m_{h_1}=m_{h_2})$.
Observe that the in-medium effects on the photon yield may come
from both the cross-section and the phase-space factor.
Consider the reaction $\pi\,\pi\,\ra\,\rho\,\gamma$. From simple kinematics 
we calculate the energy carried by the photon as 
$E_\gamma=(\sqrt{s}-m_\rho^2/\sqrt{s})/2$.  When the rho mass drops,
the possibility of getting a photon of higher energy increases for
a given centre of mass energy of the $\pi\pi$ system.  
The increase in photon emission rate with medium effects, hence, is due to 
the increase in the basic cross section $\sigma_{\pi \pi \ra \rho \gamma}$. 
The effect of the phase space factor is rather small in this case.

\begin{figure}
\centerline{\psfig{figure=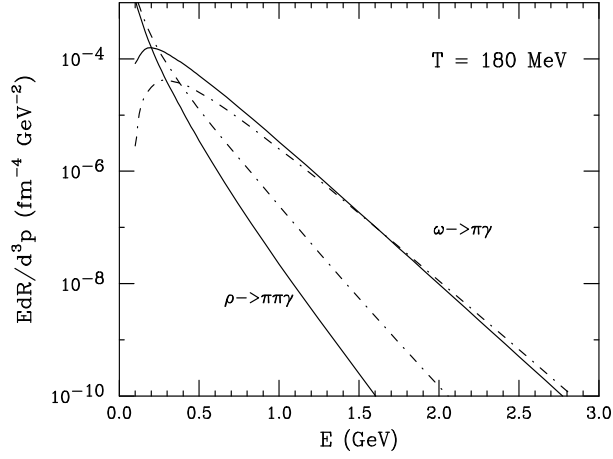,height=6cm,width=8cm}}
\caption{
Photon emission rates from $\rho \ra \pi \pi \gamma$ and $\omega \ra
\pi \gamma$ as a function of photon energy at $T$=180 MeV. 
The solid and dot-dashed lines show results
with and without in-medium effects respectively.
}
\label{dec18}
\end{figure}

In the case of the reaction $\pi\,\rho\,\ra\,\pi\gamma$, the
phase-space factor plays a very crucial role. 
At low energy ($E \leq 0.5$ GeV) the photon yield with medium
effects is found to increase by an order of magnitude. At higher
energies the enhancement is by a factor $\sim$ 2 - 3.  
At lower temperature, the medium effects on the photon
spectra should be smaller which is borne out very clearly 
from our calculations, as shown in Fig.~(\ref{phot16}).
At higher photon energies ($>0.5$ GeV) the reaction $\pi\rho\rightarrow
\pi\gamma$ dominates over $\pi\pi\rightarrow\rho\gamma$ because
for a fixed energy ($\sqrt{s}$) in the centre of mass system of the colliding
particles the production of heavier meson ($\rho$) 
leaves lesser phase-space than that for the reaction where 
 the lighter one ($\pi$) is in the final channel. In other words,
in $\pi\rho\rightarrow\pi\gamma$ the heavier meson in the incident
channel appears as a massless boson ($\gamma$) in the final 
channel making available a large amount of rest mass energy 
of the $\rho$ to the kinetic energy of the photon.  
\begin{figure}
\centerline{\psfig{figure=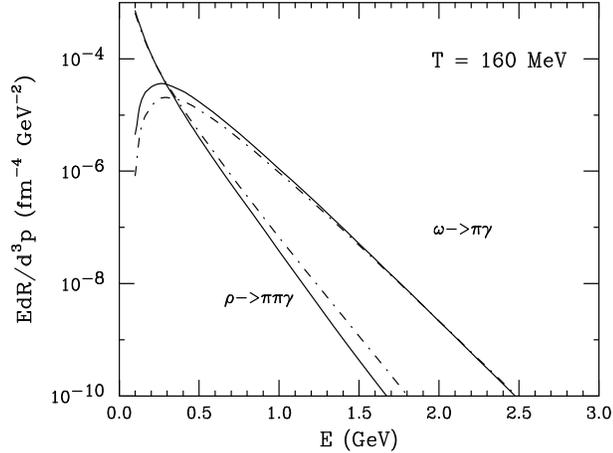,height=6cm,width=8cm}}
\caption{
Photon emission rates from $\rho \ra \pi \pi \gamma$ and $\omega \ra
\pi \gamma$ as a function of photon energy at $T$=160 MeV. 
The solid and dot-dashed lines show results
with and without in-medium effects respectively.
}
\label{dec16}
\end{figure}

The photon emission rate from the decay $\rho\,\ra\,\pi\,\pi\,\gamma$ 
in the high energy region reduces due to the decrease of the rho mass 
in the incident channel as seen in Figs.~(\ref{dec18}) and (\ref{dec16}). 
The corresponding results for $\omega$ decay are also shown. 
In the case of $\omega$ decay the total energy in the final channel is
shared by two particles whereas in the case of $\rho$ decay 
there are three particles in the final channel, therefore the former
dominates over the later at high photon energies.

\begin{figure}
\centerline{\psfig{figure=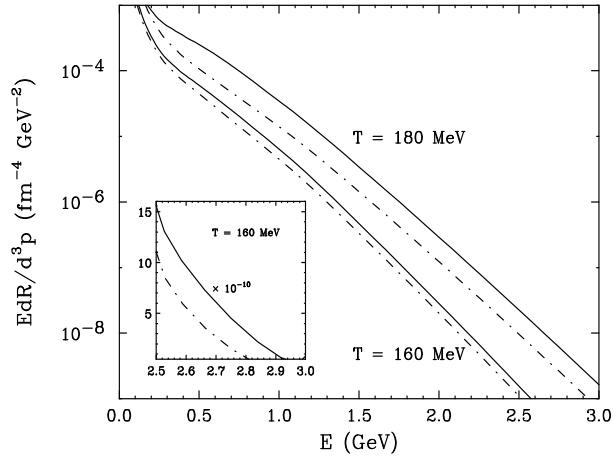,height=6cm,width=8cm}}
\caption{
Total photon emission rate from hot hadronic matter
as a function of photon energy at $T$=160, 180 MeV. 
The solid and dot-dashed lines show results
with and without in-medium effects respectively.
Inset: Total photon emission rate is plotted in linear
scale as a function of photon energy in the kinematic 
window, $E_{\gamma}=$ 2.5 to 3.0 GeV at $T=$180 MeV.
}
\label{totphot}
\end{figure}

In Fig.~(\ref{totphot}) we display the total rate of emission 
of photons from hot
hadronic gas including all hadronic reactions and decays of vector
mesons. At $T=$180 MeV, the photon emission rate with
finite temperature effects is a factor of $\sim$ 3 higher 
than the rate calculated without medium effects.
At $T=160$ MeV the medium effects are small compared to the previous
case.

\section*{V. Summary and Conclusions}

In this work we have calculated the effective masses and decay widths
of vector mesons propagating in a hot medium. We have seen that the 
mass of rho meson decreases substantially due to its interactions with
nucleonic excitations and it increases only marginally ($\sim$ 10-15 MeV) 
due to $\rho-\pi$ interactions. The overall decrease in the effective 
mass is due to fluctuations in the modified Dirac sea of nucleons with mass 
$M^{\ast}$. 
For $\omega$ mesons the in-medium contribution increases by 
70 MeV at $T=$180 MeV but the Dirac sea effect overwhelms the 
medium effect and consequently the effective omega mass drops at $T=$180
MeV by about 200 MeV from its free space value. 

We have evaluated the rate of photon emission from a hadronic gas of $\pi,
\rho, \omega$ and $\eta$ mesons. It is seen that photon rate increases by a
factor $\sim$ 3  due to the inclusion of in-medium masses at $T=$180 MeV.
We observe that the inclusion of in-medium decay width in the vector meson 
propagator has negligible effect on the photon emission rates, although 
the effect of the in-medium decay width with BE is substantial for the 
dilepton yield~\cite{Abhee}. It may be remarked that the present paper deals
with a zero net baryon density situation and is thus appropriate
for future experiments at RHIC and LHC. Inclusion of effects arising 
from finite chemical potential shall be communicated in a separate paper.

The observed photon spectra should be obtained by folding
the static rate with the expansion dynamics. 
An ideal hydrodynamical model (e.g. Bjorken hydrodynamics~\cite{Bj})
for the space time evolution does not appear to be
adequate here because of the following reason.
When the temperature of the medium is large, the hadron 
masses are reduced and the amount of mass decreased will appear
as energy of the field; as the system expands and consequently cools the 
hadrons regain their masses gradually from the field energy.
This could have non-trivial effects on the equation of state of
the system and one needs a more realistic hydrodynamic model
to handle the evolution dynamics. Hydrodynamic flow with changing hadronic 
properties is a subject of much contemporary interest and we are pursuing 
it presently~\cite{SS}.

\vskip 0.1cm
\noindent{\bf Acknowledgement:}\\
We are grateful to P. Lichard and R. Rapp for useful discussions, and
also to the referee whose comments were extremely helpful.

\section*{Appendix - I: Rho Decay Width}

\begin{figure}
\centerline{\psfig{figure=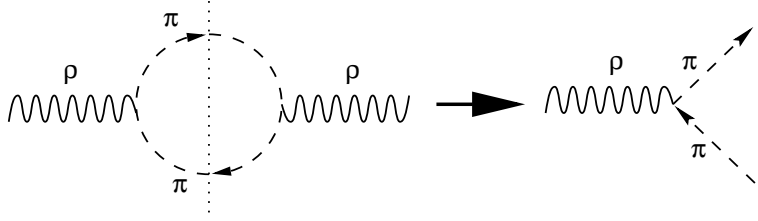,height=3cm,width=10cm}}
\caption{Decay of rho meson.}
\label{cut}
\end{figure}

We have calculated the decay width of the rho meson using the Cutkosky
rules at finite temperature (see Fig.~(\ref{cut})) which give 
a simple and systematic way to calculate the imaginary part of 
the rho self-energy. This is related to the physical decay width of the rho as
\be
{\s {Im}}\Pi(\omega) = -\omega\Gamma(\omega)
\label{impi}
\ee
The rho self-energy develops cuts along the real axis when the pions 
become on-mass shell and the discontinuity across these cuts is pure
imaginary for real $\omega$ so that we have
\be
{\s {Disc}}\Pi(\omega) = [\Pi(\omega+i\epsilon)-\Pi(\omega-i\epsilon)]
=2i{\s {Im}}\Pi(\omega)
\label{discpi}
\ee
From Eq.~(\ref{pipimn}) and the identity,
\begin{eqnarray}
\frac{1}{\beta}\sum_{n=-\infty}^{+\infty}\,f(p_0=2\pi inT)&=&
\frac{1}{2\pi i}\,\int_{-i\infty}^{i\infty}\,\frac{1}{2}[f(p_0)+f(-p_0)]dp_0
\nonumber\\
&&+\frac{1}{2\pi i}\,\int_{-i\infty+\epsilon}^{i\infty+\epsilon}\,
[f(p_0)+f(-p_0)]\frac{dp_0}{e^{\beta p_0}-1},
\end{eqnarray} 
we have, 
\begin{eqnarray}
\Pi(\omega)&=&\frac{1}{3}\Pi_{\mu}^{\mu}(\omega,{\bd k}\,\ra\,0)\nonumber\\
&=&\frac{g_{\rho \pi\pi}^2}{6\pi^2}\,\int\,\frac{p^2dp}{\omega_p}
\left[1+\frac{4p^2}{4\omega_p^2-\omega^2}\right][2n(\omega_p)+1]
\end{eqnarray} 
In the above equation $\omega=m_{\rho}^\star$ is the effective mass of 
the $\rho$ - meson.
Hence,
\begin{eqnarray}
{\s {Disc}}\Pi(\omega)&=&\frac{ig_{\rho \pi \pi}^2}{3\pi}\,\int
\frac{p^4dp}{\omega_p^2}\,[2n(\omega_p)+1]\nonumber\\
&&\times[\delta(\omega+2\omega_p)-\delta(\omega-2\omega_p)]
\label{disc1pi}
\end{eqnarray} 
From Eqs.~(\ref{discpi}) and (\ref{disc1pi}) we have
\be
{\s {Im}}\Pi(\omega) = -\frac{g_{\rho \pi \pi}^2}{48\pi \omega}
(\omega^2-4m_{\pi}^2)^{3/2}
\left[\left\{2n(\frac{\omega}{2})+1\right\}-\left\{2n(-\frac{\omega}{2})+
1\right\}\right]
\ee
We are interested in the decay width $\Gamma_{\rho\,\ra\,\pi\,\pi}$ 
and the diagram which contributes to this is shown in
Fig.~(\ref{cut}). The corresponding decay width can be read out from the 
above equation:
\be
\Gamma_{\rho\,\ra\,\pi\,\pi}(T) =  \frac{g_{\rho \pi \pi}^2}
{48\pi \omega^2}(\omega^2-4m_{\pi}^2)^{3/2}\left[2n(\frac{\omega}{2})+1\right]
\ee

\section*{Appendix - II: Invariant Amplitudes}
In this appendix we
present the invariant amplitudes for photon productions
from the Lagrangian given by Eq.~(\ref{photlag}).

\vskip .2cm
\centerline{\underline{{(1) $\pi^+(p_1)+\pi^-(p_2)\,\ra\,\rho^0(p_3)+\gamma(p_4)$}
}}
\vskip .2cm
\begin{figure}[h]
\centerline{\psfig{figure=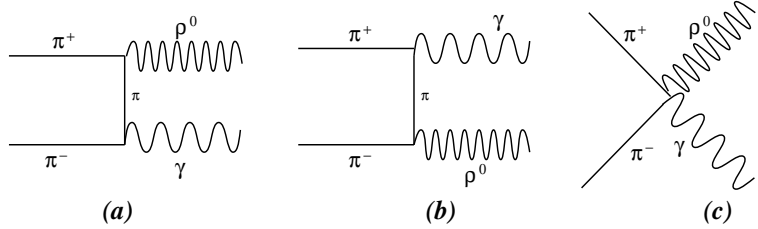,height=3cm,width=10cm}}
\caption{
Feynman diagrams for  $\pi^+\pi^-\,\ra\,\rho^0\gamma$
}
\label{reac1}
\end{figure}
\ba
\cm_a&=&-\frac{4eg_{\rpp}}{(t-m_{\pi}^2)}p_{1 \mu}p_{2 \nu} 
\ep^{\ast \mu}(p_3)\ep^{\ast \nu}(p_4)\nonumber\\
\cm_b&=&-\frac{4eg_{\rpp}}{(u-m_{\pi}^2)}p_{1 \mu}p_{2 \nu} 
\ep^{\ast \nu}(p_3)\ep^{\ast \mu}(p_4)\nonumber\\
\cm_c&=&-2eg_{\rpp}g_{\mn}\ep^{\ast \mu}(p_3)\ep^{\ast \nu}(p_4)\nonumber
\ea

\begin{eqnarray} 
\ov {\vert {\cal M}_a \vert^2}&=&\frac{16e^2g_{\rho \pi \pi}^2}
{(t-m_{\pi}^2)^2}\,
m_{\pi}^2\left[m_{\pi}^2-\frac{(m_{\pi}^2+m_{\rho}^2-t)^2}{4m_{\rho}^2}
\right]\nonumber\\
\ov {\vert {\cal M}_b \vert^2}&=&\frac{16e^2g_{\rho \pi \pi}^2}
{(u-m_{\pi}^2)^2}\,
m_{\pi}^2\left[m_{\pi}^2-\frac{(m_{\pi}^2+m_{\rho}^2-u)^2}{4m_{\rho}^2}
\right]\nonumber\\
\ov {|{\cal M}_c|^2}&=&12e^2g_{\rho \pi \pi}^2\nonumber
\end{eqnarray} 
\begin{eqnarray} 
2{\mathrm {Re}}\ov {[{\cal M}_a^{\ast}{\cal M}_b]}
&=&\frac{8e^2g_{\rho \pi \pi}^2(s-2m_{\pi}^2)}
{(t-m_{\pi}^2)(u-m_{\pi}^2)}
\left[(s-2m_{\pi}^2)-\frac{(m_{\pi}^2+m_{\rho}^2-t)(m_{\pi}^2
+m_{\rho}^2-u)}{2m_{\rho}^2}\right]\nonumber\\
2{\mathrm {Re}}\ov {[{\cal M}_a^{\ast}{\cal M}_c]}
&=&\frac{8e^2g_{\rho \pi \pi}^2}{(t-m_{\pi}^2)}
\left[(s-2m_{\pi}^2)-\frac{(m_{\pi}^2+m_{\rho}^2-u)(m_{\pi}^2+m_{\rho}^2-t)}
{2m_{\rho}^2}\right]\nonumber\\
2{\mathrm {Re}}\ov {[{\cal M}_b^{\ast}{\cal M}_c]}
&=&\frac{8e^2g_{\rho \pi \pi}^2}{(u-m_{\pi}^2)}
\left[(s-2m_{\pi}^2)-\frac{(m_{\pi}^2+m_{\rho}^2-u)(m_{\pi}^2+m_{\rho}^2-t)}
{2m_{\rho}^2}\right]\nonumber
\end{eqnarray} 
\vskip .5cm
\vskip .2cm
\centerline{\underline{{(2) $\pi^0(p_1)+\pi^{\pm}(p_2)\,\ra\,\rho^{\pm}
(p_3)+\gamma(p_4)$}
}}
\vskip .2cm
\begin{figure}[h]
\centerline{\psfig{figure=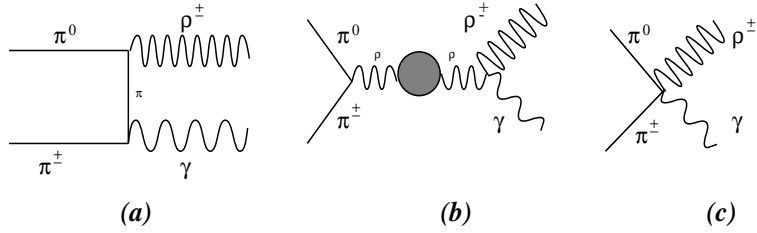,height=3cm,width=10cm}}
\caption{
Feynman diagrams for  $\pi^0\pi^{\pm}\,\ra\,\rho^{\pm}\,\gamma$
}
\label{reac2}
\end{figure}

\ba
\cm_a&=&\frac{4eg_{\rpp}}{(t-m_{\pi}^2)}p_{1 \mu}p_{2 \nu} 
\ep^{\ast \mu}(p_3)\ep^{\ast \nu}(p_4)\nonumber\\
\cm_b&=&-\frac{eg_{\rpp}}{(s-m_{\rho}^2+im_{\rho}\Gamma_{\rho})}
(p_1-p_2)^{\mu}\nonumber\\
& &\times\left[2p_{4 \nu}g_{\mu \lda}-2p_{3 \lda}g_{\mn}
+(p_3-p_4)_{\mu}g_{\nu \lda}\right]\ep^{\ast \nu}(p_3)\ep^{\ast \lda}(p_4)
\nonumber\\
\cm_c&=&eg_{\rpp}g_{\mn}\ep^{\ast \mu}(p_3)\ep^{\ast \nu}(p_4)\nonumber
\ea


\begin{eqnarray} 
\ov {\vert {\cal M}_a \vert^2}&=&\frac{16e^2g_{\rho \pi \pi}^2}{(t-m_{\pi}^2)^2}\,
m_{\pi}^2\left[m_{\pi}^2-\frac{(m_{\pi}^2+m_{\rho}^2-t)^2}{4m_{\rho}^2}
\right]\nonumber\\
\ov {|{\cal M}_b|^2}&=&\frac{e^2g_{\rho \pi \pi}^2}
{\left[(s-m_{\rho}^2)^2+m_{\rho}^2\Gamma_{\rho}^2\right]}
\left[2(t-u)^2+(4m_{\pi}^2-s)\left\{4m_{\rho}^2-\frac{(s-m_{\rho}^2)^2}
{m_{\rho}^2}\right\}\right]\nonumber\\
\ov {|{\cal M}_c|^2}&=&3e^2g_{\rho \pi \pi}^2\nonumber
\end{eqnarray} 
\begin{eqnarray} 
2{\mathrm {Re}}\ov {[{\cal M}_a^{\ast}{\cal M}_b]}
&=&\frac{4e^2g_{\rho \pi \pi}^2(s-m_{\rho}^2)}
{(t-m_{\pi}^2)[(s-m_{\rho}^2)^2+m_{\rho}^2\Gamma_{\rho}^2]}
\left[2m_{\pi}^2(t-u)-s(s-4m_{\pi}^2)\right.\nonumber\\
&&+\left.\frac{(s-4m_{\pi}^2)(s-m_{\rho}^2)(m_{\pi}^2+m_{\rho}^2-t)}
{2m_{\rho}^2}\right]\nonumber\\
2{\mathrm {Re}}\ov {[{\cal M}_a^{\ast}{\cal M}_c]}
&=&\frac{4e^2g_{\rho \pi \pi}^2}{(t-m_{\pi}^2)}
\left[(s-2m_{\pi}^2)-\frac{(m_{\pi}^2+m_{\rho}^2-u)(m_{\pi}^2+m_{\rho}^2-t)}
{2m_{\rho}^2}\right]\nonumber\\
2{\mathrm {Re}}\ov {[{\cal M}_b^{\ast}{\cal M}_c]}
&=&\frac{e^2g_{\rho \pi\pi}^2}{m_{\rho}^2}
\frac{(t-u)(5m_{\rho}^2-s)(s-m_{\rho}^2)}{[(s-m_{\rho}^2)^2+m_{\rho}^2
\Gamma_{\rho}^2]}\nonumber
\end{eqnarray} 

\vskip .2cm
\centerline{\underline{{(3) $\pi^{\pm}(p_1)+\rho^0(p_2)\,\ra\,\pi^{\pm}
(p_3)+\gamma(p_4)$}
}}
\vskip .2cm
\begin{figure}[htb]
\centerline{\psfig{figure=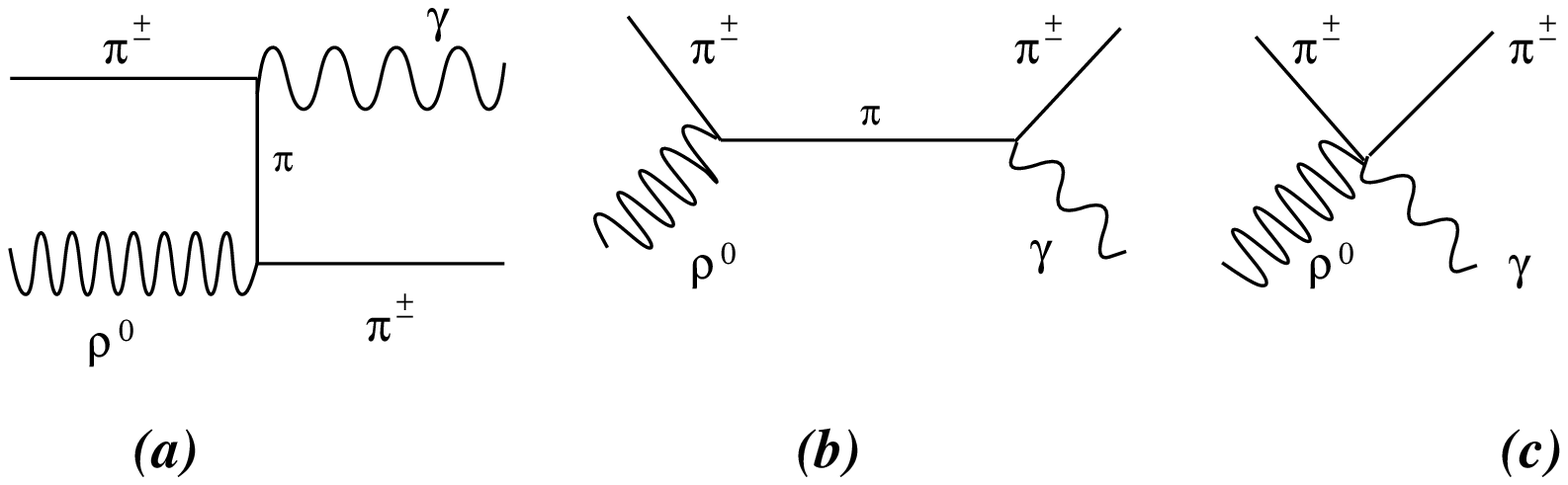,height=3cm,width=10cm}}
\caption{
Feynman diagrams for  $\pi^{\pm}\rho^0\,\ra\,\pi^{\pm}\,\gamma$
}
\label{reac3}
\end{figure}
\ba
\cm_a&=&\frac{4eg_{\rpp}}{(u-m_{\pi}^2)}p_{1 \mu}p_{3 \nu} 
\ep^{\nu}(p_2)\ep^{\ast \mu}(p_4)\nonumber\\
\cm_b&=&\frac{4eg_{\rpp}}{(s-m_{\pi}^2)}p_{1 \mu}p_{3 \nu} 
\ep^{\mu}(p_2)\ep^{\ast \nu}(p_4)\nonumber\\
\cm_c&=&-2eg_{\rpp}g_{\mn}\ep^{\mu}(p_2)\ep^{\ast \nu}(p_4)\nonumber
\ea

\begin{eqnarray} 
\ov {\vert {\cal M}_a \vert^2}&=&\frac{16e^2g_{\rho \pi \pi}^2}
{3(u-m_{\pi}^2)^2}\,
m_{\pi}^2\left[m_{\pi}^2-\frac{(m_{\pi}^2+m_{\rho}^2-u)^2}{4m_{\rho}^2}
\right]\nonumber\\
\ov {\vert {\cal M}_b \vert^2}&=&\frac{16e^2g_{\rho \pi \pi}^2}
{3(s-m_{\pi}^2)^2}\,
m_{\pi}^2\left[m_{\pi}^2-\frac{(m_{\pi}^2+m_{\rho}^2-s)^2}{4m_{\rho}^2}
\right]\nonumber\\
\ov {|{\cal M}_c|^2}&=&4e^2g_{\rho \pi \pi}^2\nonumber
\end{eqnarray} 
\begin{eqnarray} 
2{\mathrm {Re}}\ov {[{\cal M}_a^{\ast}{\cal M}_b]}
&=&\frac{4e^2g_{\rho \pi \pi}^2(t-2m_{\pi}^2)}
{3(u-m_{\pi}^2)(s-m_{\pi}^2)}
\left[(t-2m_{\pi}^2)-\frac{(m_{\pi}^2+m_{\rho}^2-s)(m_{\pi}^2
+m_{\rho}^2-u)}{2m_{\rho}^2}\right]\nonumber\\
2{\mathrm {Re}}\ov {[{\cal M}_a^{\ast}{\cal M}_c]}
&=&\frac{4e^2g_{\rho \pi \pi}^2}{3(u-m_{\pi}^2)}
\left[(t-2m_{\pi}^2)-\frac{(m_{\pi}^2+m_{\rho}^2-u)(m_{\pi}^2+m_{\rho}^2-s)}
{2m_{\rho}^2}\right]\nonumber\\
2{\mathrm {Re}}\ov {[{\cal M}_b^{\ast}{\cal M}_c]}
&=&\frac{4e^2g_{\rho \pi \pi}^2}{3(s-m_{\pi}^2)}
\left[(t-2m_{\pi}^2)-\frac{(m_{\pi}^2+m_{\rho}^2-s)(m_{\pi}^2+m_{\rho}^2-u)}
{2m_{\rho}^2}\right]\nonumber
\end{eqnarray} 

\vskip .5cm
\centerline{\underline{(4) $\pi^{\pm}(p_1)+\rho^{\mp}(p_2)\,\ra\,\pi^0(p_3)
+\gamma(p_4)$}
}
\vskip .2cm
\begin{figure}[htb]
\centerline{\psfig{figure=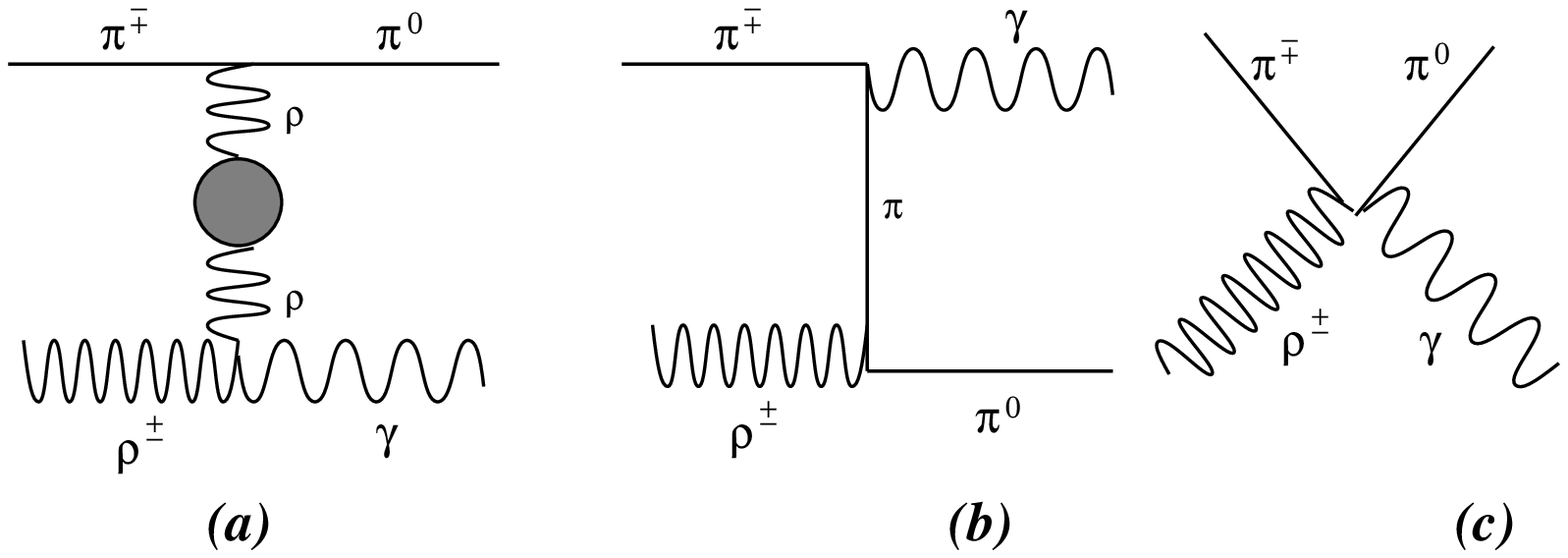,height=3cm,width=10cm}}
\caption{
Feynman diagrams for  $\pi^{\pm}\rho^{\mp}\,\ra\,\pi^0\,\gamma$
}
\label{reac4}
\end{figure}

\ba
\cm_a&=&\frac{eg_{\rpp}}{(t-m_{\rho}^2+im_{\rho}\Gamma_{\rho})}
(p_1+p_3)^{\mu}\nonumber\\
& &\times\left[2p_{4 \nu}g_{\mu \lda}+2p_{2 \lda}g_{\mn}
-(p_2+p_4)_{\mu}g_{\nu \lda}\right]\ep^{\nu}(p_2)\ep^{\ast \lda}(p_4)\nonumber\\
\cm_b&=&-\frac{4eg_{\rpp}}{(u-m_{\pi}^2)}p_{1 \mu}p_{3 \nu} 
\ep^{\nu}(p_2)\ep^{\ast \mu}(p_4)\nonumber\\
\cm_c&=&eg_{\rpp}g_{\mn}\ep^{\mu}(p_2)\ep^{\ast \nu}(p_4)\nonumber
\ea

\begin{eqnarray} 
\ov {|{\cal M}_a|^2}&=&\frac{e^2g_{\rho \pi \pi}^2}
{3\left[(t-m_{\rho}^2)^2+m_{\rho}^2\Gamma_{\rho}^2\right]}
\left[2(s-u)^2+(4m_{\pi}^2-t)\left\{4m_{\rho}^2-\frac{(t-m_{\rho}^2)^2}
{m_{\rho}^2}\right\}\right]\nonumber\\
\ov {|{\cal M}_b|^2}&=&\frac{16e^2g_{\rho \pi \pi}^2}{3(u-m_{\pi}^2)^2}\,
m_{\pi}^2\left[m_{\pi}^2-\frac{(m_{\pi}^2+m_{\rho}^2-u)^2}{4m_{\rho}^2}
\right]\nonumber\\
\ov {|{\cal M}_c|^2}&=&e^2g_{\rho \pi \pi}^2\nonumber
\end{eqnarray} 
\begin{eqnarray} 
2{\mathrm {Re}}\ov {[{\cal M}_a^{\ast}{\cal M}_b]}
&=&\frac{4e^2g_{\rho \pi \pi}^2(t-m_{\rho}^2)}
{3(u-m_{\pi}^2)[(t-m_{\rho}^2)^2+m_{\rho}^2\Gamma_{\rho}^2]}
\left[2m_{\pi}^2(u-s)-t(t-4m_{\pi}^2)\right.\nonumber\\
&&+\left.\frac{(t-4m_{\pi}^2)(t-m_{\rho}^2)(m_{\pi}^2+m_{\rho}^2-u)}
{2m_{\rho}^2}\right]\nonumber\\
2{\mathrm {Re}}\ov {[{\cal M}_a^{\ast}{\cal M}_c]}
&=&\frac{e^2g_{\rho \pi\pi}^2}{3m_{\rho}^2}
\frac{(u-s)(5m_{\rho}^2-t)(t-m_{\rho}^2)}{[(t-m_{\rho}^2)^2+m_{\rho}^2
\Gamma_{\rho}^2]}\nonumber\\
2{\mathrm {Re}}\ov {[{\cal M}_b^{\ast}{\cal M}_c]}
&=&\frac{4e^2g_{\rho \pi \pi}^2}{3(u-m_{\pi}^2)}
\left[(t-2m_{\pi}^2)-\frac{(m_{\pi}^2+m_{\rho}^2-s)(m_{\pi}^2+m_{\rho}^2-u)}
{2m_{\rho}^2}\right]\nonumber
\end{eqnarray} 

\vskip .5cm
\vskip .2cm
\centerline{\underline{(5) $\pi^0(p_1)+\rho^{\pm}(p_2)\,\ra\,\pi^{\pm}
(p_3)+\gamma(p_4)$}
}
\vskip .2cm
\begin{figure}[htb]
\centerline{\psfig{figure=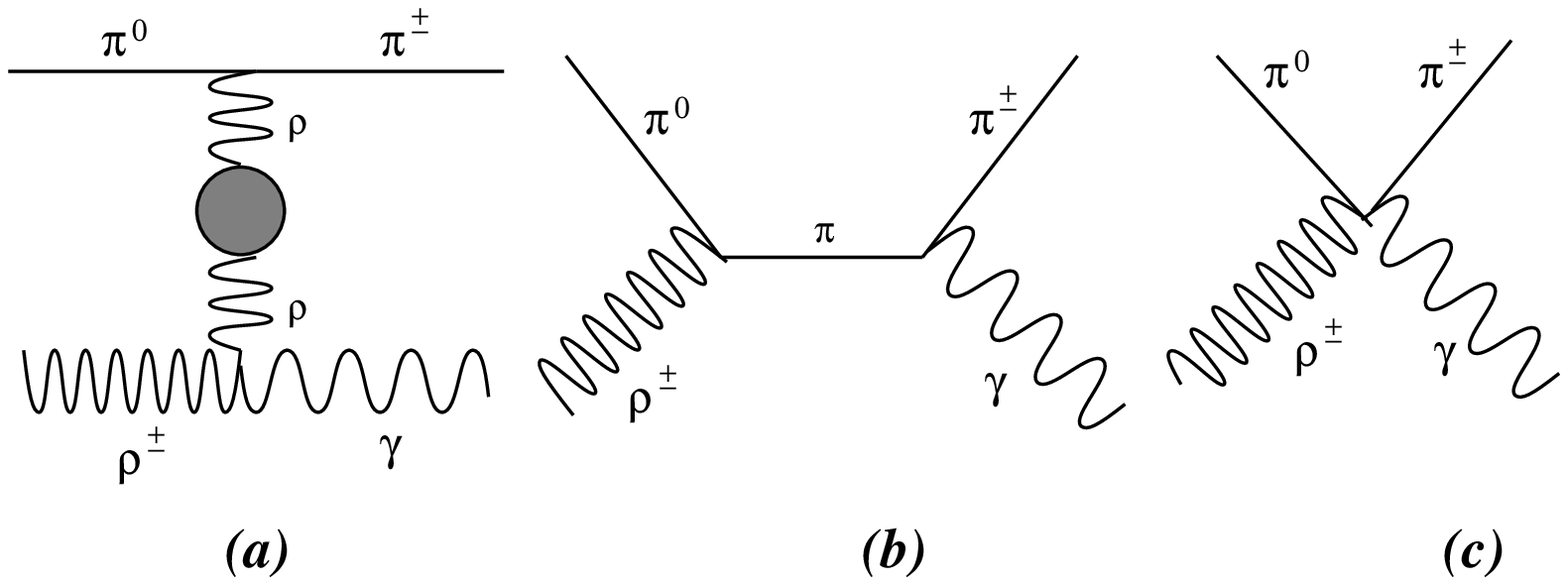,height=3cm,width=10cm}}
\caption{
Feynman diagrams for  $\pi^0\rho^{\pm}\,\ra\,\pi^{\pm}\,\gamma$
}
\label{reac5}
\end{figure}

\ba
\cm_a&=&-\frac{eg_{\rpp}}{(t-m_{\rho}^2+im_{\rho}\Gamma_{\rho})}
(p_1+p_3)^{\mu}\nonumber\\
& &\times\left[2p_{4 \nu}g_{\mu \lda}+2p_{2 \lda}g_{\mn}
-(p_2+p_4)_{\mu}g_{\nu \lda}\right]\ep^{\nu}(p_2)\ep^{\ast \lda}(p_4)\nonumber\\
\cm_b&=&-\frac{4eg_{\rpp}}{(s-m_{\pi}^2)}p_{1 \mu}p_{3 \nu} 
\ep^{\mu}(p_2)\ep^{\ast \nu}(p_4)\nonumber\\
\cm_c&=&eg_{\rpp}g_{\mn}\ep^{\mu}(p_2)\ep^{\ast \nu}(p_4)\nonumber
\ea
\begin{eqnarray} 
\ov {|{\cal M}_a|^2}&=&\frac{e^2g_{\rho \pi \pi}^2}
{3\left[(t-m_{\rho}^2)^2+m_{\rho}^2\Gamma_{\rho}^2\right]}
\left[2(s-u)^2+(4m_{\pi}^2-t)\left\{4m_{\rho}^2-\frac{(t-m_{\rho}^2)^2}
{m_{\rho}^2}\right\}\right]\nonumber\\
\ov {|{\cal M}_b|^2}&=&\frac{16e^2g_{\rho \pi \pi}^2}{3(s-m_{\pi}^2)^2}\,
m_{\pi}^2\left[m_{\pi}^2-\frac{(m_{\pi}^2+m_{\rho}^2-s)^2}{4m_{\rho}^2}
\right]\nonumber\\
\ov {|{\cal M}_c|^2}&=&e^2g_{\rho \pi \pi}^2\nonumber
\end{eqnarray} 
\begin{eqnarray} 
2{\mathrm {Re}}\ov {[{\cal M}_a^{\ast}{\cal M}_b]}
&=&\frac{4e^2g_{\rho \pi \pi}^2(t-m_{\rho}^2)}
{3(s-m_{\pi}^2)[(t-m_{\rho}^2)^2+m_{\rho}^2\Gamma_{\rho}^2]}
\left[2m_{\pi}^2(s-u)-t(t-4m_{\pi}^2)\right.\nonumber\\
&&+\left.\frac{(t-4m_{\pi}^2)(t-m_{\rho}^2)(m_{\pi}^2+m_{\rho}^2-s)}
{2m_{\rho}^2}\right]\nonumber\\
2{\mathrm {Re}}\ov {[{\cal M}_a^{\ast}{\cal M}_c]}
&=&\frac{e^2g_{\rho \pi\pi}^2}{3m_{\rho}^2}
\frac{(s-u)(5m_{\rho}^2-t)(t-m_{\rho}^2)}{[(t-m_{\rho}^2)^2+m_{\rho}^2
\Gamma_{\rho}^2]}\nonumber\\
2{\mathrm {Re}}\ov {[{\cal M}_b^{\ast}{\cal M}_c]}
&=&\frac{4e^2g_{\rho \pi \pi}^2}{3(s-m_{\pi}^2)}
\left[(t-2m_{\pi}^2)-\frac{(m_{\pi}^2+m_{\rho}^2-u)(m_{\pi}^2+m_{\rho}^2-s)}
{2m_{\rho}^2}\right]\nonumber
\end{eqnarray} 
\vskip .2cm
\centerline{\underline{(6) $\rho(p_1)\,\ra\,\pi(p_2)\,\pi(p_3)\gamma(p_4)$,
$\omega(p_1)\,\ra\,\pi(p_2)\,\gamma(p_3)$}}
\vskip .2cm
\begin{figure}
\centerline{\psfig{figure=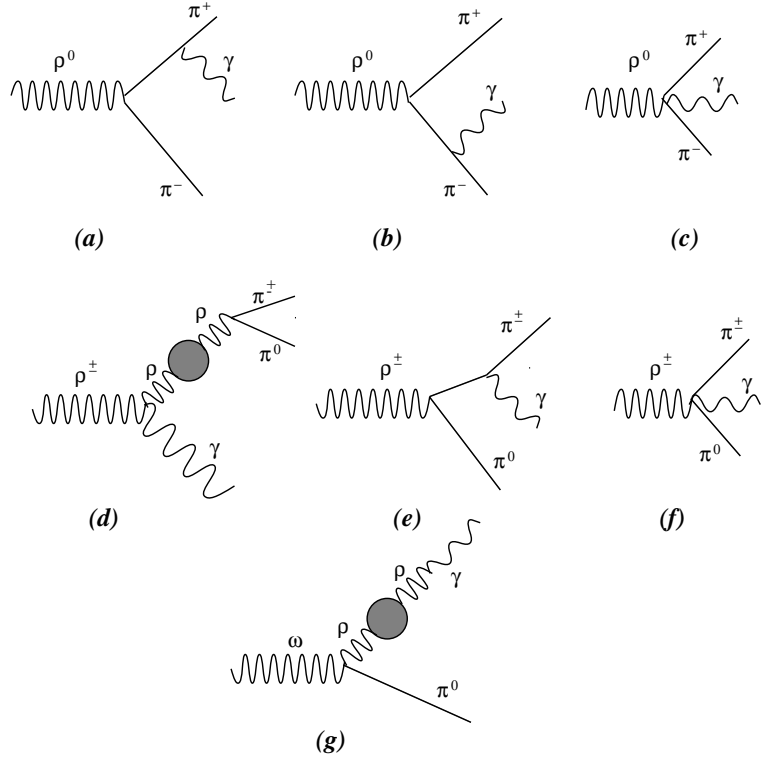,height=10cm,width=10cm}}
\caption{
Feynman diagrams for vector meson decays.
}
\label{vecdec}
\end{figure}
\ba
\cm_a&=&-\frac{4eg_{\rpp}}{(t-m_{\pi}^2)}
p_{3 \mu}p_{2 \nu}\ep^{\mu}(p_1)\ep^{\ast \nu}(p_4)\nonumber\\
\cm_b&=&-\frac{4eg_{\rpp}}{(s-m_{\pi}^2)}
p_{2 \mu}p_{3 \nu}\ep^{\mu}(p_1)\ep^{\ast \nu}(p_4)\nonumber\\
\cm_c&=&-2eg_{\rpp}g_{\mn}\ep^{\mu}(p_1)\ep^{\ast \nu}(p_4)\nonumber
\ea

\begin{eqnarray} 
\ov {\vert {\cal M}_a \vert^2}&=&\frac{16e^2g_{\rho \pi \pi}^2}
{3(t-m_{\pi}^2)^2}\,
m_{\pi}^2\left[m_{\pi}^2-\frac{(m_{\pi}^2+m_{\rho}^2-t)^2}{4m_{\rho}^2}
\right]\nonumber\\
\ov {\vert {\cal M}_b \vert^2}&=&\frac{16e^2g_{\rho \pi \pi}^2}
{3(s-m_{\pi}^2)^2}\,
m_{\pi}^2\left[m_{\pi}^2-\frac{(m_{\pi}^2+m_{\rho}^2-s)^2}{4m_{\rho}^2}
\right]\nonumber\\
\ov {|{\cal M}_c|^2}&=&4e^2g_{\rho \pi \pi}^2\nonumber
\end{eqnarray} 
\begin{eqnarray} 
2{\mathrm {Re}}\ov {[{\cal M}_a^{\ast}{\cal M}_b]}
&=&\frac{8e^2g_{\rho \pi \pi}^2(u-2m_{\pi}^2)}
{3(t-m_{\pi}^2)(s-m_{\pi}^2)}
\left[(u-2m_{\pi}^2)-\frac{(m_{\pi}^2+m_{\rho}^2-t)(m_{\pi}^2
+m_{\rho}^2-s)}{2m_{\rho}^2}\right]\nonumber\\
2{\mathrm {Re}}\ov {[{\cal M}_a^{\ast}{\cal M}_c]}
&=&\frac{8e^2g_{\rho \pi \pi}^2}{3(t-m_{\pi}^2)}
\left[(u-2m_{\pi}^2)-\frac{(m_{\pi}^2+m_{\rho}^2-s)(m_{\pi}^2+m_{\rho}^2-t)}
{2m_{\rho}^2}\right]\nonumber\\
2{\mathrm {Re}}\ov {[{\cal M}_b^{\ast}{\cal M}_c]}
&=&\frac{8e^2g_{\rho \pi \pi}^2}{3(s-m_{\pi}^2)}
\left[(u-2m_{\pi}^2)-\frac{(m_{\pi}^2+m_{\rho}^2-s)(m_{\pi}^2+m_{\rho}^2-t)}
{2m_{\rho}^2}\right]\nonumber
\end{eqnarray} 


\ba
\cm_d&=&-\frac{eg_{\rpp}}{(u-m_{\rho}^2+im_{\rho}
\Gamma_{\rho})}(p_2-p_3)^{\mu}\nonumber\\
& &\times\left[2p_{4 \nu}g_{\mu \lda}+2p_{1 \lda}g_{\mn}
-(p_1+p_4)_{\mu}g_{\nu \lda}\right]\ep^{\nu}(p_1)\ep^{\ast \lda}(p_4)\nonumber\\
\cm_e&=&\frac{4eg_{\rpp}}{(t-m_{\pi}^2)}
p_{3 \mu}p_{2 \nu}\ep^{\mu}(p_1)\ep^{\ast \nu}(p_4)\nonumber\\
\cm_f&=&eg_{\rpp}g_{\mn}\ep^{\mu}(p_1)\ep^{\ast \nu}(p_4)\nonumber
\ea
\begin{eqnarray} 
\ov {|{\cal M}_d|^2}&=&\frac{e^2g_{\rho \pi \pi}^2}
{3\left[(u-m_{\rho}^2)^2+m_{\rho}^2\Gamma_{\rho}^2\right]}
\left[2(t-s)^2+(4m_{\pi}^2-u)\left\{4m_{\rho}^2-\frac{(u-m_{\rho}^2)^2}
{m_{\rho}^2}\right\}\right]\nonumber\\
\ov {|{\cal M}_e|^2}&=&\frac{16e^2g_{\rho \pi \pi}^2}{3(t-m_{\pi}^2)^2}\,
m_{\pi}^2\left[m_{\pi}^2-\frac{(m_{\pi}^2+m_{\rho}^2-t)^2}{4m_{\rho}^2}
\right]\nonumber\\
\ov {|{\cal M}_f|^2}&=&e^2g_{\rho \pi \pi}^2\nonumber
\end{eqnarray} 
\begin{eqnarray} 
2{\mathrm {Re}}\ov {[{\cal M}_d^{\ast}{\cal M}_e]}
&=&\frac{4e^2g_{\rho \pi \pi}^2(u-m_{\rho}^2)}
{3(t-m_{\pi}^2)[(u-m_{\rho}^2)^2+m_{\rho}^2\Gamma_{\rho}^2]}
\left[2m_{\pi}^2(t-s)-u(u-4m_{\pi}^2)\right.\nonumber\\
&&+\left.\frac{(u-4m_{\pi}^2)(u-m_{\rho}^2)(m_{\pi}^2+m_{\rho}^2-t)}
{2m_{\rho}^2}\right]\nonumber\\
2{\mathrm {Re}}\ov {[{\cal M}_d^{\ast}{\cal M}_f]}
&=&\frac{e^2g_{\rho \pi\pi}^2}{3m_{\rho}^2}
\frac{(t-s)(5m_{\rho}^2-u)(u-m_{\rho}^2)}{[(u-m_{\rho}^2)^2+m_{\rho}^2
\Gamma_{\rho}^2]}\nonumber\\
2{\mathrm {Re}}\ov {[{\cal M}_e^{\ast}{\cal M}_f]}
&=&\frac{4e^2g_{\rho \pi \pi}^2}{3(t-m_{\pi}^2)}
\left[(u-2m_{\pi}^2)-\frac{(m_{\pi}^2+m_{\rho}^2-t)(m_{\pi}^2+m_{\rho}^2-s)}
{2m_{\rho}^2}\right]\nonumber
\end{eqnarray} 


$$
\cm_g=\left(\frac{eg_{\opr}m_{\rho}^2}{m_{\pi}g_{\rpp}}\right)
\ep_{\mu \nu \alpha \beta}p_1^{\mu}p_3^{\alpha}\left(\frac
{-g^{\sigma \beta}+p_3^{\sigma}p_3^{\beta}/m_{\rho}^2}{p_3^2-m_{\rho}^2
+im_{\rho}\Gamma_{\rho}}\right)\ep^{\nu}(p_1)\ep_{\sigma}^{\ast}(p_3)
$$
$$
|\cm_g|^2=\frac{2\pi\alpha}{3}\,\left(\frac{g_{\opr}}{g_{\rpp}}\right)^2
\,\frac{m_{\rho}^4}{m_{\pi}^2}\,\frac{(m_{\omega}^2-m_{\pi}^2)^2}
{[(t-m_{\rho}^2)^2+m_{\rho}^2 \Gamma_{\rho}^2]}
$$
\section*{Appendix - III}
In this appendix we define various parameters and integration limits
which appeared in Eq. (\ref{photrate}).
\begin{eqnarray}
a&=&-(s+t-m_2^2-m_3^2)^2\nonumber\\
b&=&E_1(s+t-m_2^2-m_3^2)(m_2^2-t)+E[(s+t-m_2^2-m_3^2)(s-m_1^2-m_2^2)\nonumber\\
&&-2m_1^2(m_2^2-t)]\nonumber\\
c&=&-E_1^2(m_2^2-t)^2-2E_1E[2m_2^2(s+t-m_2^2-m_3^2)-(m_2^2-t)(s-m_1^2-m_2^2)]
\nonumber\\
&&-E^2[(s-m_1^2-m_2^2)^2-4m_1^2m_2^2]-(s+t-m_2^2-m_3^2)(m_2^2-t)\nonumber\\
&&\times(s-m_1^2-m_2^2)
+m_2^2(s+t-m_2^2-m_3^2)^2+m_1^2(m_2^2-t)^2\nonumber\\
E_{1{\s {min}}}&=&\frac{(s+t-m_2^2-m_3^2)}{4E}+\frac{Em_1^2}{s+t-m_2^2-m_3^2}
\nonumber\\
E_{2{\s {min}}}&=&\frac{Em_2^2}{m_2^2-t}+\frac{m_2^2-t}{4E}\nonumber\\
E_{2{\s {max}}}&=&-\frac{b}{a}+\frac{\sqrt{b^2-ac}}{a}\nonumber
\end{eqnarray}


\end{document}